# Prediction and Control of Focal Seizure Spread: Random Walk with Restart on Heterogeneous Brain Networks

Chen Wang [a], Sida Chen [a], Liang Huang [a,b], and Lianchun Yu [a,b] *

[a] School of Physical Science and Technology, Lanzhou University, Lanzhou, 730000, China

[b] Lanzhou Center for Theoretical Physics, Key Laboratory of Theoretical Physics of Gansu Province,

Lanzhou University, Lanzhou, Gansu 730000, China

* Corresponding author: Lianchun Yu (yulch@lzu.edu.cn).

# Abstract:

Whole-brain models offer a promising method of predicting seizure spread, which is critical for successful surgery treatment of focal epilepsy. Existing methods are largely based on structural connectome, which ignores the effects of heterogeneity in regional excitability of brains. In this study, we used a whole-brain model to show that heterogeneity in nodal excitability had a significant impact on seizure propagation in the networks, and compromised the prediction accuracy with structural connections. We then addressed this problem with an algorithm based on random walk with restart on graphs. We demonstrated that by establishing a relationship between the restarting probability and the excitability for each node, this algorithm could significantly improve the seizure spread prediction accuracy in heterogeneous networks, and was more robust against the extent of heterogeneity. We also strategized surgical seizure control as a process to identify and remove the key nodes (connections) responsible for the early spread of seizures from the focal region. Compared to strategies based on structural connections, virtual surgery with a strategy based on mRWER generated outcomes with a high success rate while maintaining low damage to the brain by removing fewer anatomical connections. These findings may have potential applications in developing personalized surgery strategies for epilepsy.

**Keywords:** Focal epilepsy, Random walk, Whole-brain model, Heterogeneous networks, Virtual brain surgery

# I. Introduction

Epilepsy is a common neurological disorder characterized by unpredictable and recurrent incidents of seizures [1] that is presumed to be related to the hypersynchrony of neurons or abnormalities in neuronal network structure [2,3]. Focal seizures, which are defined as seizures that develop in a particular area (seizure focus) of the brain but can spread to other areas, are the most common type of seizure [4]. For patients with drug-resistant focal epilepsy, in addition to focal resection, transecting the fibers that are postulated to be involved in the spread of seizures is a common type of surgical treatment [5-7]. However, the success of surgery depends largely on the accurate prediction of seizure spread, which requires increasing support from quantitative analysis methods [8].

Recent research has revealed that epilepsy is a brain network disorder with widespread network effects [9-11]. For example, the association of epilepsy with abnormal functional and structural connections (SCs) among brain regions has been widely observed with advanced brain imaging techniques, such as magnetic resonance imaging (MRI) [12-16]. In addition, evidence suggests that the effects of focal epilepsy may involve areas far from the epileptic focus [17,18]. Therefore, larger-scale human brain network models constructed based on simplified local dynamics and whole-brain anatomical connections usually obtained with diffusive tensor imaging (DTI) data have become a powerful tool to study epilepsy as a brain network disorder [19]. For example, based on the conjecture that nearly every brain region can be driven out of the 'healthy' subspace to produce seizures, Jirsa et al. constructed a phenomenological model called Epileptor, which described seizure onset with saddle-node bifurcation dynamics and seizure offset with homoclinic bifurcation dynamics [20]. Using the Epileptor model to represent the local dynamics, they constructed a personalized whole-brain model to predict seizure spread that was consistent with the patient's clinical stereotactic electroencephalogram (SEEG) recording [21]. An increasing number of studies have documented the ability of personalized whole-brain network models to predict seizure spread, develop efficient surgery strategies, forecast surgical outcomes [22,23], and provide practical guidance for clinical treatment of epilepsy [22,24-26].

While structural connectivity is generally accepted as a vital determinant that influences the propagation of seizure activities in brain networks, increasing evidence suggests that regional excitability may exert substantial effects on seizure onset and propagation. For example, most antiepileptic drugs treat epilepsy by regulating cortical excitability [27]. Similarly, neurostimulation therapies such as deep brain stimulation inhibit the generation or spread of seizures by modulating neuronal network excitability [28-30]. In addition, researchers have shown that whether the type of seizure is focal or global depends on the interaction between the network structure and nodal excitability [31]. Although significant progress has been achieved in predicting and controlling seizure spread in whole-brain models, most of these strategies are based on personalized structural connectivity alone, and few researchers have considered the effects of heterogeneity in regional excitability on the prediction accuracy and control efficiency [23,24]. Obviously, given the important role of regional excitability in seizure propagation, strategies considering heterogeneity should yield more accurate prediction and efficient control of seizure spread.

Random walk simulates a particle that iteratively moves from a node to a randomly selected neighboring node [32]. It lies at the core of many algorithms to elucidate various types of structural properties of networks [33]. Among them, the random walk with restart (RWR) algorithm mimics the behavior of a particle that not only moves randomly from one node to another depending on the topological structure of the network but also restarts in the same node—called a seed. These behaviors provide a measure of the proximity or relevance between the seed and all other nodes in the network [34]. Compared with RWR, the random walk with extended restart (RWER) provides more effective relevance scores between nodes by allowing a distinct restart probability for each node [35]. The advantages of RWER algorithms in dealing with heterogeneous networks have been discussed in many studies [36,37]. In the whole-brain networks with heterogeneous excitability, through combining information from both the SCs of the network and the nodal excitability, relevance scores obtained by RWER should achieve more precisely measurement of the nodal epileptogenicity than that given by SCs alone, so to provide more accurate prediction and efficient control strategies for seizure spread.

In this study, we used the structural connectome estimated from human brain DTI data to construct whole-brain network models, and simulated the seizure spread from focal nodes. By varying the location of the focal nodes, we showed that the heterogeneity in nodal excitability exerts a substantial effect on seizure spread for most of the focal locations and subsequently jeopardize the prediction accuracy and surgery efficiency based on SCs. The relevance score of a particular node obtained by RWER algorithm proposed by Jin et al., is dependent on the restart probability of its neighboring nodes [35], whereas in our model the epileptogenicity of a node is more decided by its own excitability. To eliminate the inaccuracy due to this mismatch, we modified the original RWER model to allow the relevance score of each node to directly correlate with their restart probability. We found this modified algorithm was more precise and robust in predicting seizure spread and identifying key nodes responsible for the early stage of seizure spread. We also showed that compared with the strategy based on SCs, surgical outcomes of the strategy based on the mRWER algorithm achieved a high success rate with reduced damage to the brain. Meanwhile, the surgery efficiency was robust against the extent of heterogeneity. Finally, we also discussed several issues that may bridge the gap between this study and future clinical applications.

This paper is organized as described below. In Section II, we present the Epileptor model, which describes the nodal dynamics, the structural connectome derived from DTI data, and the approach used to build the whole-brain network model. In this section, we also present the nodal epileptogenicity measured by either SC or mRWER algorithm, and the method to evaluate the accuracy of seizure spread prediction with different measures of epileptogenicity. Section III is devoted to investigating the effect of heterogeneous excitability on seizure spread in the networks, comparing the ability of SC and mRWER epileptogenicity to predict seizure spread, and identifying key nodes responsible for the early stage of seizure spread. The efficiency of different surgical strategies is also evaluated in this section. The discussion and conclusions are presented in Sections IV and V, respectively.

# II. Models and Methods

## A. The Epileptor model

We considered networks with *n* nodes whose dynamics were described by the following version of the Epileptor model [20]:

$$\begin{aligned}
\dot{x}_1 &= y_1 - f_1(x_1, x_2) - z + I_1, \\
\dot{y}_1 &= 1 - 5(x_1)^2 - y_1, \\
\dot{z} &= \frac{1}{\tau_0}(4(x_1 - x_0) - z), \\
\dot{x}_2 &= -y_2 + x_2 - (x_2)^3 + I_2 + 0.002g(x_1) - 0.3(z - 3.5), \\
\dot{y}_2 &= \frac{1}{\tau_2}(-y_2 + f_2(x_1, x_2)),
\end{aligned} \quad (1)$$

where

$$f_1(x_1, x_2) = \begin{cases} x_1^3 - 3x_1^2, & if\ x_1 < 0, \\ (x_2 - 0.6(z-4)^2)x_1, & if\ x_1 \geq 0, \end{cases} \quad (2)$$

$$f_2(x_1, x_2) = \begin{cases} 0, & if\ x_2 < -0.25, \\ 6(x_2 + 0.25), & if\ x_2 \geq -0.25, \end{cases} \quad (3)$$

and

$$g(x_1) = \int_{t_0}^{t} e^{-\gamma(t-\tau)} x_1(\tau) dt. \quad (4)$$

The parameters used in this study was listed in Table I. Epileptor is a phenomenological model in which the physiological mechanism of seizure generation is replaced by the equivalent dynamic mechanism [20]. It comprises one subsystem (subsystem 1) with two state variables ($x_1$ & $y_1$) responsible for generating fast discharges and another subsystem (subsystem 2) with two state variables ($x_2$ & $y_2$) generating sharp-wave events. The function $g(x_1)$ is a low-pass filtered excitatory coupling from subsystem 1 to 2 to generate the SWE and interictal spikes. Note that the integral coupling function $g(x_1)$ can be rewritten as an ordinary differential equation (see Ref. [20] for details). The function $f_1(x_1, x_2)$ is a linear inhibitory coupling from subsystem 2 to 1. The two subsystems are linked to the so-called permittivity variable *z*, which evolves on a very slow timescale. The permittivity variable *z* is presumed to be related to the excitability, and dictates how close the system is to the seizure threshold [38,39]. Therefore, in this study we adjusted $x_0$ to determine whether the node is a 'healthy' node or not, and to control the excitability of the healthy node. In detail, there exists a threshold value of $x_{0c}$ ($x_{0c} = -2.05$ in the current setting of parameters). If $x_0 < x_{0c}$, the node is a healthy node because the system would stay at a stable fixed point, and no seizure occurs. If we increased $x_0$ so that $x_0 > x_{0c}$, the system would undergo a transition to the ictal periods through the saddle-node bifurcation, the node becomes a focal node and generates seizure activities constantly (Fig. 1(b)). It is obvious that the excitability of a healthy node is decided by the distance between $x_0$ and $x_{0c}$. If $x_0$ is closer to $x_{0c}$, the node would more likely to cross the seizure threshold under external interventions, and become a seizure-recruited node.

## B. Anatomical connectivity matrix

Brain regions exchange information by sending and receiving signals through white matter fibers which also form the seizure propagation pathways in brains of epilepsy patients. DTI is capable of providing a non-invasive estimation of the whole brain structural connections by tracking these fibers. The fiber count derived from DTI data has been confirmed to provide a realistic estimate of white matter pathway projection strength, and has been widely used to construct whole brain models [40]. In this study, we used diffusion MRI images of 284 healthy subjects from the HCP 1200-subject release to determine the strength of internode structural connections. Detailed information on these data is available at https://www.humanconnectome.org. The DTI data were processed with DSI Studio (http://dsi-studio.labsolver.org). The detailed processing procedure could be found in Ref.[41] . In short, the whole brain was parcellated into $n = 116$ regions with the Anatomical Automatic Labeling Atlas (https://www.gin.cnrs.fr/en/tools/aal/) [42]. The SC matrix $S$ was calculated using the count of the connecting tracks and was normalized by dividing by its maximum.

## C. The whole-brain model

The whole-brain model was constructed by coupling the 116 Epileptors with the SC matrix $S$ in the following diffusive form [43]:

$$
\begin{aligned}
\dot{x}_{1,i} &= y_{1,i} - f_1(x_{1,i}, x_{2,i}) - z_i + I_1, \\
\dot{y}_{1,i} &= 1 - 5(x_{1,i})^2 - y_{1,i}, \\
\dot{z}_i &= \frac{1}{\tau_0}\left(4(x_{1,i} - x_{0,i}) - z_i - \sum_{j=1}^{n} S_{ij} \cdot (x_{1,j} - x_{1,i})\right), \\
\dot{x}_{2,i} &= -y_{2,i} + x_{2,i} - (x_{2,i})^3 + I_2 + 0.002 g(x_{1,i}) - 0.3(z_i - 3.5), \\
\dot{y}_{2,i} &= \frac{1}{\tau_2}\left(-y_{2,i} + f_2(x_{1,i}, x_{2,i})\right),
\end{aligned}
\qquad (5)
$$

where subscript $i$ of each variable indicates the $i$-th node. $S_{ij}$ is the element of the $i$-th row and $j$-th column of the normalized anatomical connectivity matrix **S**, i.e., the SC strength between brain regions $i$ and $j$ derived from DTI data. This kind of slow coupling via permittivity variable $z$ represents the extracellular and intracellular effects of local and distant discharges involved in the spread of the seizure, and has been shown to play determinant factor for seizure recruitment in partial epilepsy [25]. It is noted that this is different from fast couplings through synapses or gap junctions [44].

In this study, to test the generality of out method, we created virtual brain models with different location of focal nodes, which were randomly selected with equal probability from all the 116 nodes in the network. We did not consider the case for multiple focal nodes so each brain model only had one focal node. For a particular model with the $i$-th node as the focal node, we let $x_{0,i} > x_{0c}$ to simulate the seizure onset on the $i$-th node, and let other nodes be in the healthy state (i.e., $x_{0,j} < x_{0c}, j = 1,2,\ldots,n, j \neq i$ ). With this arrangement, if we chose node B as the focal node and let other nodes in the network remain healthy state, the seizure originating from node B would cause seizure activity at other nodes (for example, node A) with a time delay (Fig. 1(d)) if the coupling between them was sufficiently large. Otherwise, the seizure would be confined to the focal node and would not spread to node A (Fig. 1(c)).

The heterogeneity in nodal excitability was simulated for the nodes in the network other than the focal node:

$$
x_{0,i} \sim N(\mu, \sigma^2), x_{0,i} < x_{0c}, for\ i = 1,2,\ldots,n, i \neq focal, \qquad (6)
$$

where $N(\mu, \sigma^2)$ is a normal distribution with mean $\mu$ and variance $\sigma^2$. The network would be heterogeneous if $\sigma > 0$ and homogeneous if $\sigma = 0$. In subsequent simulations, we set $\mu = -2.12$, unless specified otherwise. The current study was limited to the case in which only one focal node was present in the network. So the standard deviation $\sigma$, which indicates the extent of heterogeneity, should not be too large. Otherwise, seizures could be generated on multiple nodes at the same time, which is more analogous to the generalized epilepsy. In the simulations, the Euler-Maruyama method with an integration step of 0.05 was used to numerically solve the whole-brain network models.

Table I. Parameters of the whole-brain models used in this study.

| | | |
|---|---|---|
| $x_{0,focal}$ | Excitability for the focal nodes | U(-0.9, -1)* |
| $x_{0,i}$ | Excitability for the healthy nodes | ** |
| $I_1$ | Passive current of subsystem 1 | 3.1 |
| $I_2$ | Passive current of subsystem 2 | 0.45 |
| $\tau_0$ | Characteristic time scale of the permittivity variable | 2857 |
| $\tau_2$ | Characteristic time scale of subsystem 2 | 10 |
| $\mu$ | Mean excitability of the healthy nodes | -2.12 |
| $\sigma$ | Extent of heterogeneity in nodal excitability | [0, 0.04] |
| $S$ | The structural connectivity matrix | ** |
| $\gamma$ | Time constant in function g(x) | 0.01 |

* U (a, b) stands for uniform distribution ranging from a to b.
** explained in the context.

## D. Measuring nodal epileptogenicity with SC/mRWER

Knowledge of the nodal epileptogenicity, i.e., the susceptibility of each node in the network to the seizure, is important to predict seizure spread from a particular focal node. In our work, high epileptogenicity of a node implies that this node is easily recruited by seizure activity through inputs from focal nodes. The SC epileptogenicity of a node considers the contribution of SC strength between this node and the focal node to its seizure onset. Therefore, the SC epileptogenicity of a node $i$ can be measured by $S_{ij}$, the SC strength between node $i$ and the focal node $j$. In this study, we also introduced the mRWER epileptogenicity, which combines the contribution of both SC strength and the nodal excitability. In the following part of this section, we described the mRWER algorithm which was modified from RWER algorithm, and the way to link this algorithm to our whole-brain model so that the mRWER epileptogenicity could be calculated.

As explained in the Introduction section, for heterogeneous whole-brain networks, in principle we can establish a link between the restarting probability of the particles and the excitability for each node, and use RWER algorithm that combines information from both SCs and the nodal excitability, to have more precisely measurement of the nodal epileptogenicity (with the stationary distribution of particles on each node) than that given by SCs alone. However, the relevance score of a particular node obtained by the RWER is more dependent on the restart probability of its neighboring nodes rather than the restart probability of itself [35]. Therefore, we modified the original RWER model to allow the relevance score of each node to directly correlate with their own restart probability. As shown in Fig. 2(a), in the original RWER, for a particular node $j$, the particles on the other nodes would restart to the seed node $k$ with their corresponding restart probability (e.g., node $i$ with restart probability $c_i$) and walk to node $j$ with probability $(1 - c_i)A_{ij}$. This process leads to the recursive equation for RWER in the following form:

$$r_j^{t+1} = \sum_i (1 - c_i) A_{ij} r_i^t + \left(\sum_i c_i r_i^t\right) q_j, \qquad (7)$$

where $r_i^t$ is the RWER probability $r_i$ of node $i$ at time t. $A_{ij}$ is the element of the weighted adjacency matrix, $c_i \in [0,1]$ is the restart probability of node $i$, and $q_j = 1$ if node $j$ is a seed node; otherwise, $q_j = 0$. In the modified RWER, for a particular node $j$, the particles on the other nodes (e.g., node $i$) would first walk to node $j$ with probability $A_{ij}$ and then restart to the seed node $k$ with probability $c_j$. The recursive equation for mRWER is then written as follows:

$$r_j^{t+1} = (1 - c_j) \sum_i A_{ij} r_i^t + \left(\sum_i (\sum_k c_k A_{ik}) r_i^t\right) q_j. \qquad (8)$$

In Eq. (8), the first term on the right side corresponds to the random walk process, and the second term corresponds to the restart process. Equation (8) can be expressed in its matrix form:

$$\boldsymbol{r} = (\boldsymbol{I} - \text{diag}(\boldsymbol{c})) \boldsymbol{A}^T \boldsymbol{r} + ((\boldsymbol{Ac})^T \boldsymbol{r}) \boldsymbol{q}, \qquad (9)$$

or the closed form:

$$r = (I - B)^{-1}q, \tag{10}$$

where $B = (I - \text{diag}(c))A^T + q(Ac - 1)^T$. In this study, the stationary solution $r$ was calculated directly from Eq. (10).

For the mRWER, particles tend to stay at nodes with a lower restart probability, rather than nodes whose neighbors have a lower restart probability. If the weighted adjacency matrix A was decided by SC matrix $S$, the seed node was taken as the focal node, and the restart probability of other nodes was set to be inversely proportional to their excitability; the stationary probability $r_i$ for a particular node $i$ can be used as a measure of the epileptogenicity for node $i$. The mRWER epileptogenicity is expected to perform better than SC epileptogenicity in heterogeneous networks because it utilizes additional information from nodal excitability.

Notably, matrix $A$ in mRWER must be row-normalized to ensure that the iteration in Eq. (8) converges [35]. Meanwhile, matrix A must be proportional to the SC matrix $S$ such that mRWER is able to generate predictions about our whole-brain model. For these purposes, matrix A in our work was determined from the SC matrix S as follows: 1) Normalization of the SC matrix: $S'_{ij} = S_{ij}/\max(D_{ii})$, where $D_{ii} = \sum_j S_{ij}$ and 2) Construction of row-normalized matrix A according to $S'_{ij}$: $A_{ij} = S'_{ij}, i \neq j$; $A_{ii} = 1 - \sum_j S'_{ij}$. The corresponding restart probability $c_i$ for node $i$ in the RWER was determined by its excitability parameter $x_{0,i}$ as follows:

$$c_i = sigmoid\left(-b(x'_{0,i} - x_{0c})\right), for\ i = 1,2,\dots,n, i \neq focal, \tag{11}$$

where $b$ is a positive scale factor allowing mRWER to match seizure propagation in the network, and $b = 22$. The sigmoid function $sigmoid(x) = \frac{1}{1+e^{-x}}$ ensures that the mapped values of $c_i$ are in the interval of [0, 1], and $x'_{0,i}$ is the effective heterogeneity after node $i$ with $x_{0,i}$ is introduced into our whole-brain network with diffusive coupling,

$$x'_{0,i} = x_{0,i} + 0.1 \sum_{j=1}^{n} S_{ij}(x_{0,j} - x_{0,i}), for\ i = 1,2,\dots,n, i \neq focal. \tag{12}$$

By incorporating SC matrix $S$ and nodal excitability $x_{0,i}$ ($i = 1,2,\dots,n, i \neq focal$) of our whole-brain model into the mRWER algorithm, the stationary solution $r$ would be a measure of epileptogenicity for all the healthy nodes in the network with a given the focal node. However, it is impossible to compare the epileptogenicity measured for networks with different focal node, because the stationary solution $r$ obtained with mRWER algorithm was a normalized vector, and it did not reflect the overall effects on epileptogenicity of the network brought by different focal node through SCs. Therefore, to compare the nodal epileptogenicity of networks with different choices of focal nodes, the mRWER epileptogenicity for node $i$ was determined as the value $r_i$ of the steady-state distribution $r$ multiplied by the degree of the focal node in the SC matrix, i.e., the summarized SC strength between the focal node and other nodes. For simplicity, the mRWER epileptogenicity of the focal node was always set to 0.

## E. Measuring the accuracy of seizure spread prediction

One aim of this study is to investigate whether the nodal epileptogenicity measured by either SC or mRWER could accurately predict the seizure spread patterns characterized by nodal onset delay time. In this study, the prediction accuracy was evaluated by the normalized discounted cumulative gain (nDCG), which calculated the similarity between nodal rankings based on their epileptogenicity and nodal rankings based on their onset delay time [45].

In detail, the discounted cumulative gain calculates the gain for each recruited node based on its position in the nodal onset delay list $O$ that sorted from largest to smallest, and multiplied by the discount based on its position in the nodal epileptogenicity list $E$ that sorted from largest to smallest. Supposing $o_i$ is the position of node $i$ in the nodal onset delay list $O$, the gain of node $i$ was calculated by $2^{o_i} - 1$, which implied that nodes with shorter onset delay time would have higher gain. For simplicity, the position $o_i$ was set to zero if node $i$ was not recruited or it was the focal node, so that their gain was zero. With this arrangement, $o_i$ ranges from 0 to $m$, where $m$ is the total number of recruited nodes. Supposing $e_i$ is the position (from 1 to 116) of recruited node $i$ in the nodal epileptogenicity list $E$, the discount of node $i$ was calculated by $\frac{1}{\log_2(e_i+1)}$, which implied that nodes with higher epileptogenicity would have larger discount. Then the discounted gains were summarized across all the nodes in the network to provide a measure of similarity between the above two sorted lists:

$$DCG_m = \sum_{i=1}^{116} \frac{2^{o_i}-1}{\log_2(e_i+1)}. \tag{13}$$

It is noted that the value of $DCG_m$ depends on the total number of recruited nodes $m$, which varies in simulations. To compare the $DCG_m$ values with different $m$, we normalized the $DCG_m$ values obtained from a particular $m$ by $DCG_m^{max}$, the maximal value that measured similarity between a list with length $m$ and a list with same length but reverse order:

$$nDCG = \frac{DCG_m}{DCG_m^{max}} = \sum_{i=1}^{116} \frac{2^{o_i}-1}{\log_2(e_i+1)} / \sum_{i=1}^{m} \frac{2^i-1}{\log_2(m-i+2)}. \tag{14}$$

Theoretically, the values of $nDCG$ lies in the range between 0 and 1.

## III. Results

### A. Prediction of seizure spread in the heterogeneous whole-brain networks

We first investigated how the heterogeneous distribution of nodal excitability influences seizure spread in the whole-brain networks. For this purpose, we simulated seizure spread in both homogeneous and heterogeneous whole-brain networks, and compared the corresponding propagation patterns characterized by sequences of nodal onset delay times. Figure 3(a) presents an example of seizure spread in homogeneous networks. The seizure activity originating from the focal node would spread to the other nodes with different delay times. To investigate whether seizure spread patterns could be predicted by SCs, we measured the SC epileptogenicity of each node with the SC strength between them and the focal node, and tested whether there was a significant correlation between the SC epileptogenicity of each node and their onset delay time. Since the early stage of seizure spread in the homogeneous network is decided only by the SCs between the recruited nodes and the focal node, we observed a significant correlation between the onset delay time of these nodes and their SC epileptogenicity (Fig. 3(b)). Next, we set the focal node as the seed node and obtained the mRWER epileptogenicity for each node by solving Eq. (10) for the stationary solutions of the particle distribution probability across all other nodes. Since the excitability was the same for all the nodes in the homogeneous networks, the mRWER epileptogenicity was decided only by matrix **A**, i.e., the SCs. Therefore, similar to the SC epileptogenicity, there existed significant correlation between mRWER epileptogenicity and onset delay time (Fig. 3(c)).

For the heterogeneous networks, we observed a significantly altered seizure propagation pattern though the seizure was initiated at the same focal node (Fig. 3(d)). Because the seizure spread in heterogeneous networks was not determined by SCs alone but was a result of interaction between SCs and nodal excitability, the correlation between onset delay and SC epileptogenicity was no longer significant (Fig. 3(e)). However, in this case, mRWER epileptogenicity, which considered not only the effects of SCs but also the nodal excitability on the seizure spread, still maintained a significant correlation with the onset delay time (Fig. 3(f)).

Next, we used the nDCG to measure the accuracy of seizure spread prediction with SC and mRWER epileptogenicity for different choices of focal node. As demonstrated in Fig. 4(a), the prediction accuracy of both methods varied as different nodes were chosen as the focal nodes. The prediction accuracy of SC epileptogenicity decreased significantly when heterogeneity was introduced (the third panel from top to bottom in Fig. 4(a)). However, the prediction accuracy of mRWER epileptogenicity remained approximately the same as that of SC epileptogenicity for the homogeneous networks (Fig. 4(b)), and was significantly higher than that of SC epileptogenicity for the heterogeneous networks (Fig. 4(c)). Furthermore, we calculated the mean $nDCG$ value (averaged across the realizations with different choices of focal nodes) for networks with different extent of heterogeneity $\sigma$ defined in Eq. (6) to test how the extent of heterogeneity affects the prediction accuracy of these two methods in heterogeneous networks. As illustrated in Fig. 4(d), as the extent of heterogeneity increased, the averaged prediction accuracy decreased significantly for SC epileptogenicity (the blue dots in the violin plots), especially for some network configuration involving focal nodes and nodal excitability (the lower thin dashed lines indicating the quartiles on the left side of the violin plots). However, the averaged prediction accuracy with mRWER epileptogenicity was robust to the extent of heterogeneity (the orange dots in the violin plots), with less of a decrease for different network configuration involving focal nodes and nodal excitability (the lower thin dashed lines indicating the quartiles on the right side of the violin plots). Therefore, we concluded that for the heterogeneous brain networks, the seizure spread could be predicted more accurately and robustly by mRWER epileptogenicity rather than SC epileptogenicity.

## B. Identification of key nodes for seizure spread

Based on the discussion above, nodes with direct SCs to the focal node are the nodes firstly recruited by seizure activity in the early stage of seizure spread. Among these nodes, there exists a minimal set of nodes that play a crucial role in seizure spread. Disconnecting the SCs between these nodes and the focal node will absolutely block the seizure spread. We defined this minimal set of nodes with respect to the focal node as the key nodes to block the seizure spread. The key nodes were identified in our whole-brain model by progressively removing the SCs between other nodes and the focal node according to their onset delay times. For example, as shown in Figure 5(a), in the original network (network 1), we first ran the simulation of seizure spread from the focal node (node 2) to find that node 58 has the shortest onset delay time. We then removed the SCs between this node and the focal node to generate network 2. Again, we ran the simulation of seizure spread in network 2 and found that it had a different seizure propagation pattern and obviously prolonged onset delay time than network 1 (middle panel in Fig. 5 (a)). Then, we removed the SCs between the focal node and the node with the shortest onset delay time (node 8 in this case) to generate network 3. By simulating seizure spread in network 3, we did not detect seizure activity that spread from the focal node (right panel in Fig. 5 (a)). Therefore, node 58 and node 8 were determined as the key nodes responsible for seizure spread in network 1.

Although ideally the key nodes for seizure spread could be identified using the approach described above through simulations of whole-brain networks, this approach is usually not practical in the clinic. On the other hand, the key nodes may be identified from the difference in onset delay time of each node in the original networks. In the aforementioned example, the key nodes have significant shorter onset delay times in the original network than other nodes (non-key nodes, Fig. 5(b)). However, this property is not always true because we also observed cases in which the key nodes were those with longer onset delay times than other nodes (as observed for the outliers in Fig. 5(c) for the homogeneous network and Fig. 5(d) for the heterogeneous network). Anyway, the average onset delay time of the key nodes is always significantly shorter than that of other nodes, regardless of whether the network is homogeneous or heterogeneous (Fig. 5(b) and (c)).

The shorter average onset delay time for the key nodes compared with other nodes implies that the key nodes can be distinguished from other nodes through their SC/mRWER epileptogenicity. To this end, we evaluated the performance of SC/mRWER epileptogenicity in the classification of the key nodes by constructing receiver operating characteristic (ROC) curves. For a particular network, we calculated the corresponding SC/mRWER epileptogenicity for each node and chose the nodes with SC/mRWER epileptogenicity larger than a threshold as the predicted key nodes. We then obtained the ROC curve by calculating the true positive rate (the ratio of the number of correctly predicted key nodes to the number of key nodes that were identified using the aforementioned progressive cutting procedure) and the false positive rate (the ratio of the number of falsely reported key nodes to the total number of non-key nodes) for different thresholds. Particularly, we excluded the nodes that had no or very weak structural connections to the focal node (less than 0.05 in this study) before classification because these nodes were unlikely to be recruited directly by seizure activity from the focal node. We performed the classification with the same threshold for SC/mRWER epileptogenicity on a set of networks with different choices of focal nodes. Figure 6 shows the classification performance of SC and mRWER epileptogenicity for networks with different extent of heterogeneity. The SC epileptogenicity yields the best classification performance in homogeneous networks (ROC curve corresponding to $\sigma = 0$ in Fig. 6(a)). However, the significantly dropped ROC curves as $\sigma$ increases imply the classification performance of SC epileptogenicity is jeopardized by heterogeneity (Fig. 6(a)). On the contrary, the ROC curves for mRWER epileptogenicity keep the same for different values of $\sigma$, implying its classification performance is unaffected by heterogeneity (Fig. 6(b)). As $\sigma$ is increased from 0 to 0.04, classification performance measured by the area under the ROC curve drops from 0.979 to 0.919 for SC epileptogenicity, but it keep almost unchanged (from 0.983 to 0.980 ) for mRWER epileptogenicity. These results suggest that the mRWER epileptogenicity outperformed SC epileptogenicity in the robust classification of key nodes that are responsible for seizure spread in heterogeneous networks.

## C. The efficiency of virtual surgery

It is seen that the seizure spread might be suppressed in patients with focal seizures by disconnecting the SCs between focal nodes and the key nodes classified by SC/RWER epileptogenicity. Therefore, it is interesting to investigation how the surgical outcomes depend on different classification methods and network heterogeneity. First, we described an example of virtual surgery on a virtual patient for whom node 37 is the focal node in his/her personalized whole-brain network model. Fig. 7(a) depicts the results of classification for key nodes with both SC and mRWER epileptogenicity, supposing the brain network of this patient was homogeneous. Four nodes (nodes 39, 41, 77, and 89) were identified as key nodes with a threshold of 0.19 for SC epileptogenicity (blue filled circles that were larger than others in Fig. 7(a)). The same four nodes were identified as key nodes with a threshold of 0.05 for RWER epileptogenicity (red filled squares that have larger sizes than others in Fig. 7(a)). Simulation of the whole-brain network model after removing the SCs between these four key nodes and the focal node showed that the seizure spread was blocked (Fig. 7(b)). Unlike the homogeneous network, the different classification methods yielded different numbers of key nodes in the heterogeneous network with the same threshold as in the homogeneous network. As shown in Fig. 7(c), nodes 39, 41, 77, and 89 were again identified as key nodes with SC epileptogenicity (threshold: 0.19, blue filled circles with a larger size than others in Fig. 7(c)), whereas only node 55 and node 89 were identified as key nodes with mRWER epileptogenicity (threshold: 0.05, red filled squares with a larger size than others in Fig. 7(c)). We performed virtual surgery on this heterogeneous whole-brain network model and found that either removing the SCs between the four key nodes identified by SC epileptogenicity and the focal node (Fig. 7(d)) or removing the SCs between the two key nodes identified by mRWER epileptogenicity and the focal node (Fig. 7(e)) can block the seizure spread. However, removing the SCs between the two nodes with the highest SC epileptogenicity (i.e., node 77 and node 89) and the focal node failed to block the seizure spread (Fig. 7(f)). Therefore, although seizure spread in heterogeneous networks could be controlled by both surgical strategies, the strategy based on SC epileptogenicity caused more damage to the brain networks.

Next, to validate our conclusions for arbitrarily chosen focal nodes, we tested the surgical outcomes in a group of patients with different locations of the seizure focus. The focal node of each patient was selected randomly with equal probability from the 116 nodes in the network. We performed the aforementioned virtual surgery procedure to evaluate the success rate and surgical damage for different surgical strategies. Here, the success rate refers to the proportion of patients whose seizure spread is successfully blocked after surgery, and damage rate refers to the average ratio of the number of SCs that must be removed to the maximal possible number of SCs between the focal node and other nodes (i.e., 116) to successfully control seizure spread. We used the success-damage curve to describe the surgical outcomes of different surgical strategies. Similar to the ROC curve analysis, these two measures critically depend on the threshold of SC or mRWER epileptogenicity. If the threshold is too large, no node would be classified as a key node so that both the success rate and damage rate would be zero. If the threshold is too low, all nodes would be classified as key nodes so that the damage rate is that all connections must be removed, though in this case the success rate would be 100%.

Fig. 8 (a) and 8 (b) show the success-damage curve for virtual surgery on a group of patients with strategies based on SC/mRWER epileptogenicity. If the surgical strategy is based on SC epileptogenicity, the curves drop significantly as the extent of heterogeneity increased in networks (Fig. 8 (a)). However, if the surgical strategy is based on mRWER epileptogenicity, the curves remained almost the same (Fig. 8 (b)). We also obtained the ideal surgical outcomes by identifying and removing the key nodes with the procedures described in Fig. 5 (a). It is noted that neither strategy achieved ideal surgical outcomes (markers at the top-left of the plots, enlarged in the inserts for better view). We then measured the surgical efficiency by calculating the areas under the success-damage curves. A larger area under the success-damage curve indicates a higher success rate with lower damage rate and thus corresponded to higher surgical efficiency. As shown in Fig. 8(c), the surgical efficiency for the strategy based on SC epileptogenicity decreased significantly as the extent of heterogeneity increased, whereas the surgical efficiency for the strategy based on mRWER epileptogenicity was more robust against changes in the extent of heterogeneity. On the other hand, the damage rate corresponding to the 95% success rate (marked with horizontal dashed lines in Fig. 8 (a) and (b)) caused by surgery with mRWER epileptogenicity was limited to around 0.04, which meant averagely 4 to 5 SCs must be removed for each patient to block the seizure spread. It changed slightly as the extent of heterogeneity increased and remained at almost the same level as in the homogeneous networks. However, the damage caused by surgery with SC epileptogenicity increased quickly as the extent of heterogeneity increased, and exceeded twice that of homogeneous networks when $\sigma$ was increased to 0.04 (Fig. 8 (d)). These results imply that surgical strategy based on mRWER epileptogenicity might achieve higher success rate with lower damage to the brain, and it is quite robust against changes in heterogeneity.

## IV. Discussion

Accurate prediction and efficient control of seizure spread are vital for the clinical treatment of epilepsy. In the present study, to achieve more accurate prediction and efficient control of seizure spread in the heterogeneous whole-brain networks, instead of using nodal epileptogenicity that derived from SCs alone, we proposed mRWER algorithm, a random-walk based measure of the nodal epileptogenicity that combines the information of both SCs and nodal excitability. We demonstrated that the mRWER epileptogenicity was superior to SC epileptogenicity in predicting seizure spread and classifying key nodes responsible for seizure spread. Furthermore, the outcomes of virtual surgery with the strategy based on mRWER epileptogenicity achieved a high success rate while maintaining rather low damage (i.e., requiring fewer SCs to be removed to block seizure spread). More importantly, while the performance of SC epileptogenicity deteriorated as the extent of network heterogeneity increased, mRWER epileptogenicity exhibited better and more robust performance.

Using advanced personalized whole-brain modeling techniques, accurate prediction and efficient control of focal seizure spread becomes possible [22,24-26]. To improve the accuracy of prediction, it is vital to incorporate into these models more biological factors that have substantial influences on seizure spread. Besides SCs, heterogeneity in regional excitability is the prominent factors that have been argued to affect the seizure spread. First, although many previous studies used identical parameters for nodal dynamics to construct whole-brain network models [23,24,46], an increasing number of study showed that including heterogeneity in the model would yield a more realistic simulation of network dynamics in the brain [47,48]. Recent studies have shown that incorporating heterogeneity inferred from the MRI-derived T1- to T2-weighted (T1w/T2w) index generates more realistic resting-state functional connectivity measured by functional MRI (fMRI)[47,48]. Second, as we argued in the Introduction section, the critical role of regional excitability in seizure onset and propagation has been gradually revealed [27,31]. Third, various techniques have been developed to infer regional cortical excitability, such as Bayesian inference based on SEEG recording [21,49,50] or noninvasive transcranial magnetic stimulation (TMS) [51]. However, few studies have investigated the effects of heterogeneity in regional excitability on seizure spread and subsequent prediction and control strategies [22,25,26]. Our work therefore filled this gap by firstly determining the effect of heterogeneity on seizure spread in a whole-brain model and then proposed a random-walk based method that yielded more accurate prediction and efficient control of seizure spread than methods based on SCs alone.

In the present study, we showed that the nodal epileptogenicity measured by mRWER is more accurate than that of SC in predicting seizure spread in heterogeneous networks. The underlying mechanism can be understood as follows. The onset of seizures can be considered a threshold crossing event [20]. Therefore, seizure onset at one particular node depends on two factors. The first one is the summarized inputs from other nodes via SCs. The larger this summarized input is, the easier subsystem 1 reaches the threshold. The second one is excitability of this node which is decided by the distance between $x_0$ and $x_{0c}$. With the same amount of inputs, the node with higher excitability (short distance between $x_0$ and $x_{0c}$) would be easier to generate seizure activity by crossing the threshold. The prediction of seizure spread with SC epileptogenicity works well for homogeneous networks because only the first factor need be considered in this case. It is not surprise to see that its performance becomes worse once the heterogeneity is included in the model. On the contrary, the mRWER epileptogenicity not only includes the first factor through the SC matrix but also includes the second factor by linking the nodal restart probability with their excitability.

Lack of clinical trials to support out results is one major limitation of this study. However, we would like to note several issues that would help to bridge the gap between our work and future clinical applications. First, the SC matrix used in this study was derived from a DTI dataset of 286 healthy subjects. Although this matrix would not affect the general conclusions of our study, this SC matrix is not the same as that used for clinical purposes. In the clinic, since the focal node and SCs are patient-specific, the SC matrix of patients would generate more accurate predictions through the use of the personal SC matrix [22,24]. Second, in this study, regional excitability was set randomly, whereas recent studies on brain imaging techniques suggested that although large intersubject variability in regional excitability occurs, certain patterns of regional excitability exist across brain regions [47,48]. Thus, those details could be incorporated into the patient's personalized brain models to yield better predictions. Third, in the present study, we removed all connections from the focal node to the key nodes. However, in practice, removing a portion of the connections is also an optional strategy to control seizures. Therefore, it is interesting to investigate whether the mRWER algorithm still produces results in this case.

# V. Conclusions

In conclusion, brain networks are heterogeneous, and seizure propagation in brain networks is significantly affected by heterogeneity in regional excitability. Therefore, the accuracy of the prediction for seizure propagation based on structural connections alone is reduced by heterogeneity. However, the mRWER algorithm proposed in this study mimics the essential dynamics of seizure onset in heterogeneous networks, thus providing better predictions of seizure propagation. The outcomes of virtual surgery suggest that the strategy based on the mRWER algorithm would have a high success rate while maintaining a low level of brain damage by removing fewer structural connections (i.e., fewer lesions in white matter fibers).

These findings may have potential applications in the field of personalized brain modeling and personalized surgery strategy for patients with epilepsy.

# Acknowledgement

This study was funded by the Fundamental Research Funds for the Central Universities (Grant No. lzujbky-2021-62), and the National Natural Science Foundation of China (NSFC No. 12047501).

# Figures and Legends:
## Figure 1:

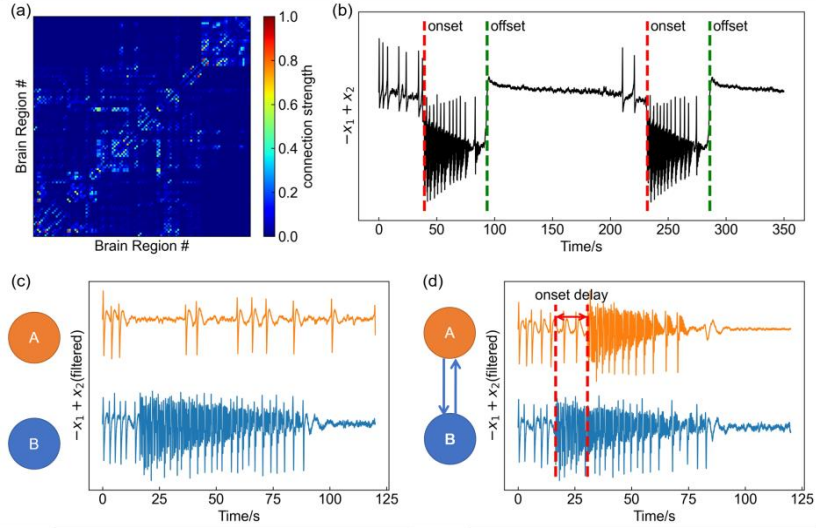

**Figure 1 Construction of the whole-brain model for seizure onset and spread.** (a) The normalized anatomical connectivity matrix derived from DTI data. (b) Seizure activity generated by a single Epileptor model. In this plot, the signal is derived from the directed sum of discharges, $-x_1 + x_2$, as the Epileptor model suggested. (c) Filtered signals of two independent nodes. The excitability of healthy node A is lower than the threshold, i.e., $x_{0,A} = -2.1$. The excitability of epileptic node B is higher than the threshold, i.e., $x_{0,B} = -2$. (d) Filtered signal of two nodes in the network with the same excitability in (c). Due to coupling, healthy node A is recruited after seizure onset at node B (the focal node). In (b) - (d), the 256 time steps were set to 1 s to match the actual seizure frequency. In (c) and (d), the original signal, i.e., $-x_1 + x_2$, was filtered by a 5th-order Butterworth bandpass filter with cutoff frequencies of 0.16 and 97 Hz at -3 dB [52].

## Figure 2:

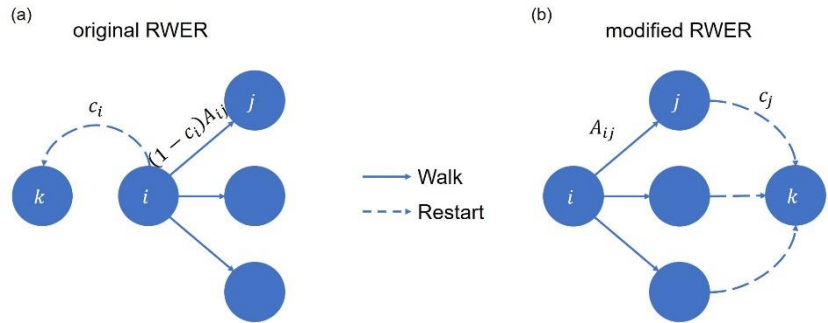

**Figure 2 Schematic diagram of the original and modified RWER algorithms.** (a) In the original RWER, for node $j$, the particles at the other node, e.g., node $i$, would restart to the seed node $k$ with a probability $c_i$ and walk to node $j$ with probability $(1 - c_i)A_{ij}$. (b) In the modified RWER, the particles at the other nodes (e.g., node $i$) would first walk to node $j$ with probability $A_{ij}$ and then restart to seed node $k$ with probability $c_j$.

**Figure 3:**

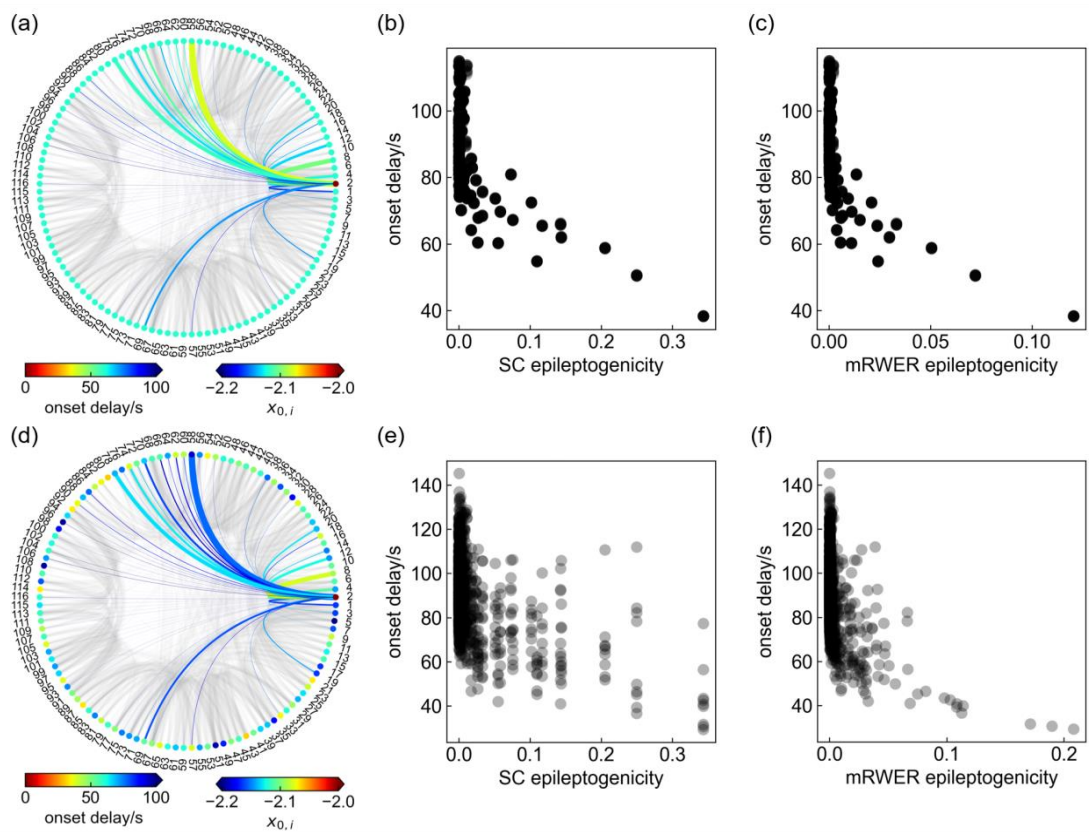

**Figure 3 Seizure spread in the homogeneous and heterogeneous whole-brain networks with arbitrarily selected focal node (in this case, node 2).** (a) Seizure spread in a homogeneous network ($\mu = -2.12$, $\sigma = 0$). In the plot, each dot represents a node of the network. The color of the dots represents the excitability of each node (horizontal color bar at the bottom right). The gray lines represent the SCs among nodes derived from the DTI data. The width of the gray lines represents the SC strength. The colored lines represent the onset delay time of nodes (horizontal color bar at the bottom left). The focal node is indicated by the filled brown dot. (b) and (c) The correlations between the nodal onset delay time and SC/mRWER epileptogenicity for seizure spread in homogeneous networks. The dots are collected from 10 simulations of the same network. (d) The same as in (a) but for a heterogeneous network ($\mu = -2.12$, $\sigma = 0.04$). (e) and (f) The correlations between the nodal onset delay time and SC/mRWER epileptogenicity for seizure spread in heterogeneous networks. The dots are collected from the simulation of 10 networks as in (d) but with different allocation of nodal excitability. In (b), (c), (e), and (f), the overlapped dots are marked with darker colors.

**Figure 4:**

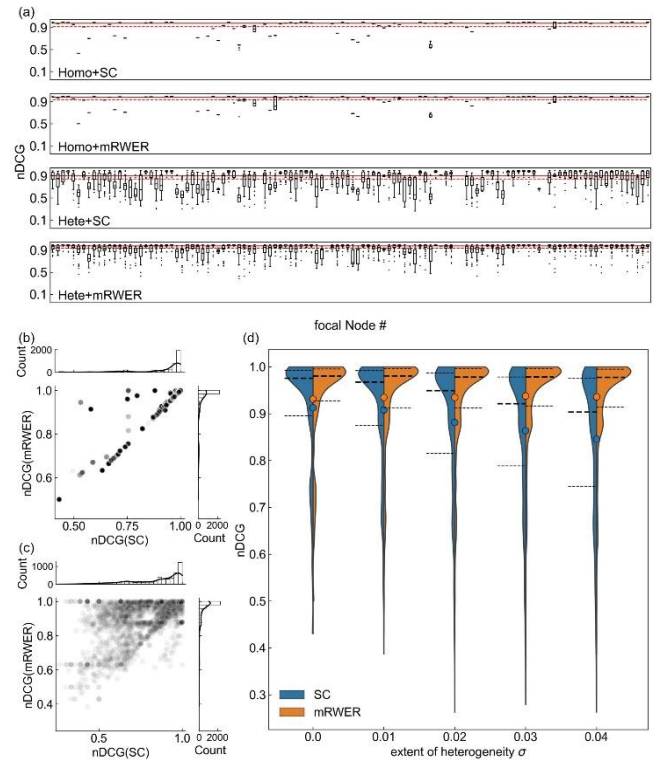

**Figure 4 The prediction accuracy of SC and mRWER epileptogenicity measured with nDCG for homogeneous and heterogeneous networks with different choices of focal node.** (a) The prediction accuracy of SC and mRWER epileptogenicity measured with nDCG for networks with different location of focal node. From top to bottom, Homo/Hete: homogeneous ($\mu = -2.12, \sigma = 0$)/heterogeneous ($\mu = -2.12, \sigma = 0.04$) networks. The red dashed lines represent the $nDCG$ value averaged across the 116 possible locations of focal nodes, which are 0.9127, 0.9309, 0.8455, and 0.9361, respectively, from top to bottom. The red solid lines are the medians of $nDCG$ values, which are 0.9754, 0.9802, 0.9032, and 0.9774, respectively, from top to bottom. (b) and (c) Comparison of the prediction ability of SC and mRWER epileptogenicity in the homogeneous (b) and heterogeneous (c) networks. The grayscale of the dots represents the number of overlapping dots. The darker the dots are, the more overlapping dots are observed at this position. (d) The dependence of the prediction accuracy of SC and mRWER epileptogenicity on the extent of heterogeneity (i.e., $\sigma$) in the networks. Here, the $nDCG$ values were calculated from the results of 4800 realizations with randomly selected focal node in each realization. Thick dashed lines in the plot are the medians, and thin dashed lines are the quartiles. The dots indicate the mean values of $nDCG$.

**Figure 5:**

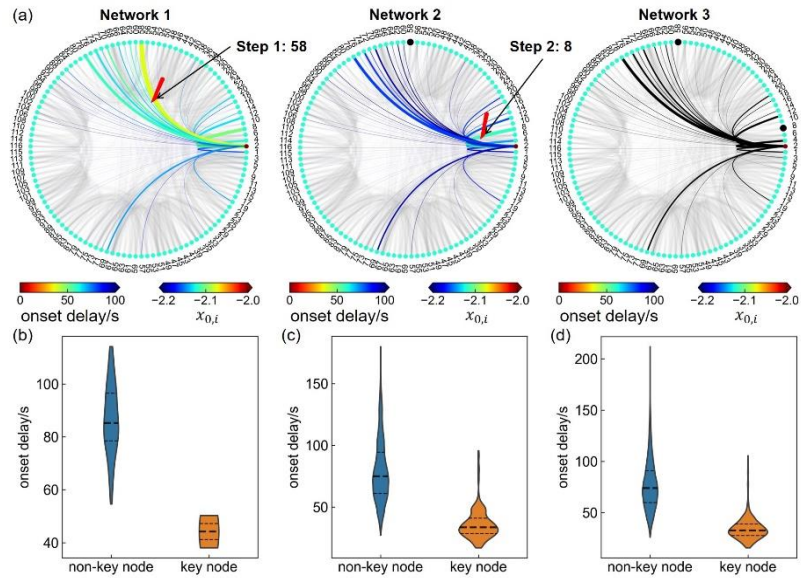

**Figure 5 Identification of the key nodes for seizure spread by progressively removing SCs between nodes with the shortest onset delay time and the focal node.** (a) An example of identifying key nodes by progressively disconnecting the SCs between nodes with the shortest delay time and the focal node. Left panel: the original network (focal node: node 2). After simulation of seizure spread in this network, the onset delay time of each node was recorded, and node 58 was determined as the key node because it had the shortest onset delay time. Middle panel: the network with disconnected SCs between node 58 and node 2. After computer simulation of seizure spread in this network, the onset delay time of each node was recorded again; node 8 was determined as the key node in this case because it had the shortest onset delay time. Right panel: the network with disconnected SCs between node 58 and node 2, as well as node 8 and node 2. Computer simulation of this network did not found any healthy node that was recruited by seizure activity (black lines represent the infinitely long onset delay time of nodes that are connected to the focal node). (b) Comparison of the onset delay time in the original network for key nodes and non-key nodes, for example, in (a). (c) Comparison of the onset delay time in the original network for key nodes and non-key nodes for homogeneous networks ($\mu = -2.12, \sigma = 0$, 800 realizations). (d) Comparison of the onset delay time in the original network for key nodes and non-key nodes for heterogeneous networks ($\mu = -2.12, \sigma = 0.04$, 800 realizations). In (b)-(d), the thick dashed lines indicate the medians, and the thin dashed lines indicate the quartiles.

**Figure 6:**

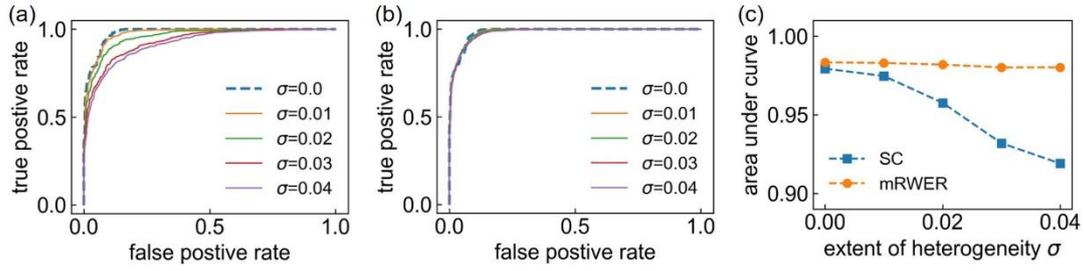

**Figure 6 Performances of key nodes classification for networks with different extents of heterogeneity.** (a) ROC curves for classification with SC epileptogenicity. (b) ROC curves for classification with mRWER epileptogenicity. (c) The classification performance measured by the area under the ROC curves for different methods.

**Figure 7:**

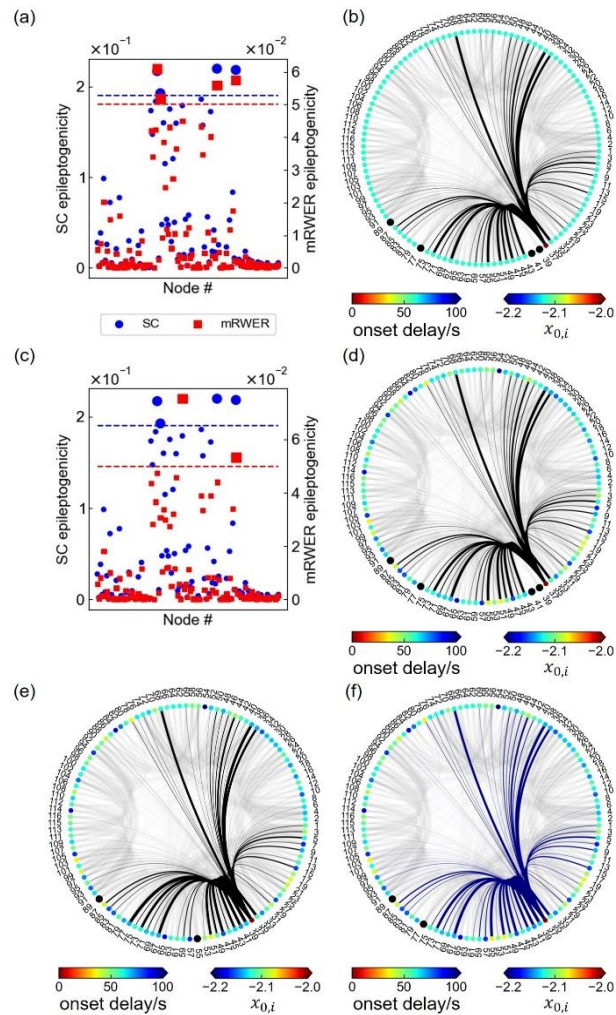

**Figure 7 Examples of virtual surgery on virtual patients using different strategies.** (a) The classification of key nodes with both SC and mRWER epileptogenicity in virtual patients whose network is homogeneous ($\mu = -2.12, \sigma = 0$) and node 37 is the focal node. The dashed lines represent the threshold to classify key nodes (blue for SCs and red for mRWER). The identified key nodes are marked with the same icons but are larger in size. In this case, the same four nodes (nodes 41, 39, 77, and 89) were identified as key nodes using different methods. (b) The seizure activity was blocked after virtual surgery was performed to remove SCs between the four nodes identified in (a) and the focal node. The plot is the same as above, but the four identified key nodes are marked with black dots. Black lines represent the infinitely long onset delay time of nodes that are connected to the focal node. (c) The same as in (a), but the network is heterogeneous ($\mu = -2.12, \sigma = 0.03$). In this case, four nodes (nodes 41, 39, 77, and 89) were identified as key nodes by SC epileptogenicity, but only two nodes (nodes 55 and 89) were identified as key nodes by mRWER epileptogenicity. (d) The seizure activity was blocked in a heterogeneous network after virtual surgery was performed to remove SCs between four nodes identified by SC epileptogenicity in (c) and the focal node. The plot is the same as in (b). (e) The seizure activity was blocked in a heterogeneous network after virtual surgery was performed to remove SCs between the two nodes identified by mRWER epileptogenicity in (c) and the focal node. The plot is the same as in (b). (f) Seizure activity was not blocked in a heterogeneous network after virtual surgery was performed to remove SCs between two of the four nodes (nodes 77 and 89) identified by SC epileptogenicity in (c) and the focal node. The plot is the same as in (b).

**Figure 8:**

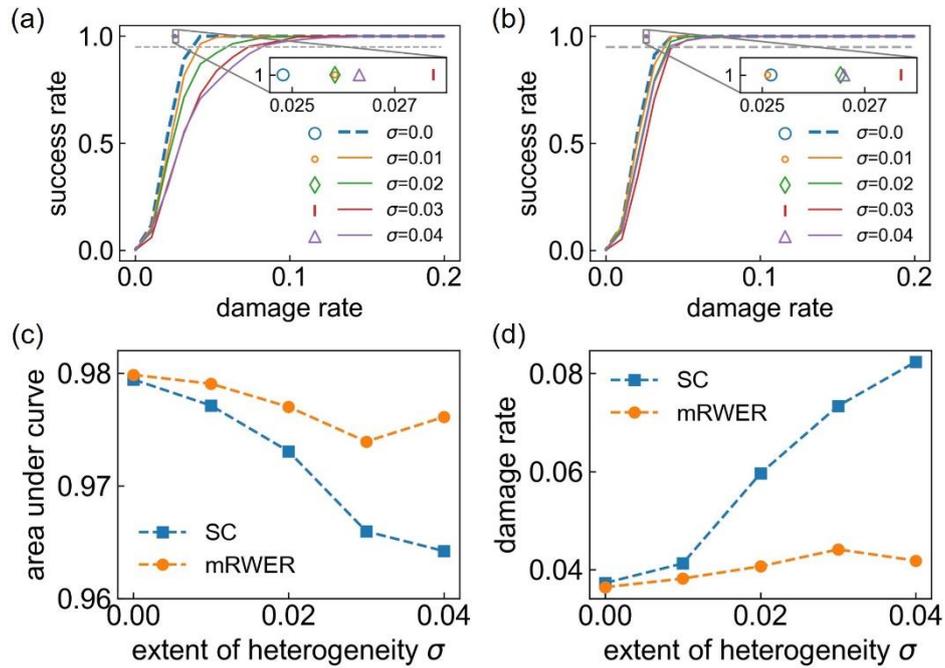

**Figure 8 The efficiency of virtual surgery based on different strategies.** (a) The success-damage curves of the surgical strategy based on SC epileptogenicity with different extents of heterogeneity. Insert: The ideal success rate vs. damage rate of surgery obtained by identifying the key nodes with the procedures described in Fig. 5(a). The gray dashed line indicates the 95% success rate. (b) The same as in (a), but for mRWER epileptogenicity. (c) The dependence of surgical efficiency evaluated based on the area under the success-damage curves for the extent of heterogeneity of different surgical strategies. (d) The dependence of damage rate at the 95% success rate on the extent of heterogeneity of different surgical strategies. The results were calculated with 800 virtual patients for each value of the extent of heterogeneity.

**30 / 30**